\documentclass[epj]{webofc}
\usepackage{amsmath}
\usepackage{bm}
\usepackage[varg]{txfonts}   % Web of Conferences font
\usepackage{pifont}% http://ctan.org/pkg/pifont
\newcommand{\cmark}{\ding{51}}%
\newcommand{\xmark}{\ding{55}}%
\wocname{EPJ Web of Conferences}
\woctitle{CONF12}
\begin{document}
\selectlanguage{english}
\title{Combination of measurements and the BLUE method}
%
% subtitle (optional, strongly discouraged)
%
%%%\subtitle{Do you have a subtitle?\\ If so, write it here}

\author{Luca Lista\inst{1}\fnsep\thanks{\email{luca.lista@na.infn.it}}}

\institute{INFN - Istituto Nazionale di Fisica Nucleare - Sezione di Napoli
}

\abstract{%
  The most accurate method to combine measurements from different experiments is to build a combined likelihood function and use it to perform the desired inference. This is not always possible for various reasons, hence approximate methods are often convenient. Among those, the best linear unbiased estimator (BLUE) is the most popular, allowing to take into account individual uncertainties and their correlations. The method is unbiased by construction if the true uncertainties and their correlations are known, but it may exhibit a bias if uncertainty estimates are used in place of the true ones, in particular if those estimated uncertainties depend on measured values. In those cases, an iterative application of the BLUE method may reduce the bias of the combined measurement.
}
\maketitle
\section{Measurement combination}

The most rigorous and accurate method to combine measurements from different
experiments relies on the combination of the individual likelihood functions
that have been used for each measurements. Imagine that
an experiment $A$ provides a measured data sample $\vec{x}_A$,
whose likelihood function, characterized by a set of parameters $\vec{\theta}_A$, is
\begin{equation}
  L(\vec{x}_A; \vec{\theta}_A)\,,
\end{equation}
and an experiment $B$ provides a sample $\vec{x}_B$,
whose likelihood function, characterized by a set of parameters $\vec{\theta}_B$, is
\begin{equation}
  L(\vec{x}_B; \vec{\theta}_B)\,.
\end{equation}
The parameter sets $\vec{\theta}_A$ and $\vec{\theta}_B$ contain some experiment-specific
parameters and some parameters common to both experiments. Among the latter, there are
physical parameters of interest, nuisance parameters related to common source of systematic uncertainties,
such as theory uncertainties, accelerator-specific component of luminosity uncertainty, etc.
The global likelihood function, if $A$ and $B$ are independent experiments, is:
\begin{equation}
  L(\vec{x}_A,\vec{x}_B; \vec{\theta}) = L(\vec{x}_A; \vec{\theta}_A) \cdot  L(\vec{x}_B; \vec{\theta}_B)\,,
\end{equation}
where $\vec{\theta}$ is the set of all common and non-common parameters.
Any statistical method can be applied at this level to determine a combined measurement, upper
limit and/or significance: a Bayesian or frequentist inference, profile likelihood,
modified frequentist upper limit, etc.

An example of such approach is provided by the combination of measurements of
single top-quark production in the $s$-channel at the Tevatron performed by
the CDF and D0 experiments~\cite{CDFs1, CDFs2, D0s}.

CDF and D0 measurements of single-top production cross section were combined using
as input the binned distributions of multivariate discriminator outputs for each individual measurement.
Each bin in each data sample was used in a Bayesian analysis, assuming a Poisson distribution,
to extract a central value of the cross-section estimate.
A likelihood-ratio analysis using asymptotic formulae~\cite{asymptotic} was
instead used to
determine the combined significance of the observed $s$-channel signal.

\section{Approximate approaches: the BLUE method}

In many cases, unless experiments, or even analysis groups within the same experiment, agree in advance,
individual likelihood functions may not be available, or are available under different software frameworks, etc.
Approximate methods can be used to combine the individual results, which are usually provided in terms of a central
value and an uncertainty:
\begin{eqnarray}
\theta & = & \hat{\theta}_1 \pm \sigma_1\label{eq:th1}\,,\\
\theta & = & \hat{\theta}_2 \pm \sigma_2\label{eq:th2}\,.
\end{eqnarray}
Correlation between uncertainties,
which is related to the  non-diagonal elements of the covariance matrix $\mathbf{V}$
\begin{equation}
\rho = \frac{V_{12}}{\sigma_1 \sigma_2}\,,
\end{equation}
must be properly taken into account.

The most popular method to combine correlated measurements
is the best linear unbiased estimate (BLUE). The method
was formulated initially in
the '30s~\cite{BLUE0} and proposed in high-energy physics in the '80s~\cite{BLUE}.
By definition, it is the unbiased linear estimator that provides the smallest possible variance
assuming the true uncertainties and their correlation are known. The estimator is 
equivalent to a $\chi^2$ minimization which, for Gaussian distributions,
is also equivalent to a maximum-likelihood estimate.

Given two measurements, as in eq.~(\ref{eq:th1}) and~(\ref{eq:th2}),
the BLUE estimate is given by the following linear combination of the individual measurements:
\begin{equation}
  \hat{\theta} = \frac{\hat{\theta}_1(\sigma_2^2 -\rho\sigma_1\sigma_2) + \hat{\theta}_2(\sigma_1^2-\rho\sigma_1\sigma_2)
  }{\sigma_1^2-2\rho\sigma_1\sigma_2+\sigma_2^2}
  \label{eq:blue}
\end{equation}
with variance:
\begin{equation}
  \sigma_{\hat{\theta}}^2 = \frac{\sigma_1^2\sigma_2^2(1-\rho^2)}{\sigma_1^2-2\rho\sigma_1\sigma_2+\sigma_2^2}\,.
  \label{eq:blueerr}
\end{equation}
Unlike the usual weighted average, for some values of the correlation coefficient $\rho$, the
coefficients of the linear combination that appear in eq.~(\ref{eq:blue}) may 
be negative.

More in general, for $n$ measurements $\hat{\theta}_1\pm\sigma_1, \cdots, \hat{\theta}_n\pm\sigma_n$,
with a covariance matrix $\mathbf{V}$, the BLUE combination in eq.~(\ref{eq:blue}) can be generalized as:

\begin{equation}
  \hat{\theta} = \sum_{i=1}^n w_i \hat{\theta}_i\,,
  \label{eq:bluen}
\end{equation}
with variance:
\begin{equation}
  \sigma_{\hat{\theta}}^2 = \vec{w}^{\mathrm{T}}\,\mathbf{V}\,\vec{w}\,,
\end{equation}
The weights in eq.~(\ref{eq:bluen}) can be computed as:
\begin{equation}
  \vec{w} = \frac{\mathbf{V}^{-1}\vec{u}}
  {\vec{u}^{\mathrm{T}}\,\mathbf{V}\,\vec{u}}
\end{equation}
where $\vec{u} = (1,\cdots,1)$ is the vector with all elements equal to unity.
The normalization condition for weights $\vec{w}$  holds:
\begin{equation}
  \sum_{i=1}^nw_i = 1\,.
\end{equation}

In case the weights $w_i$ are positive, an interpretation of the BLUE combination
in eq.~(\ref{eq:blue}) in term of weighted average is possible by introducing
the ``common error''~\cite{Greenlee}, defined as:
\begin{equation}
  \sigma_{\mathrm{c}} = \rho\sigma_1\sigma_2\,.
\end{equation}
The two measurements in eq.~(\ref{eq:th1}) and~(\ref{eq:th2})
can be rewritten with uncertainties given by the sum in quadrature of fully uncorrelated contributions
$\sigma_1^\prime$ and $\sigma_2^\prime$ and a 100\% correlated contribution $\sigma_{\mathrm{C}}$:
\begin{eqnarray}
\theta & = & \hat{\theta}_1 \pm \sigma_1^\prime \pm \sigma_{\mathrm{C}} \label{eq:thx1}\,,\\
\theta & = & \hat{\theta}_2 \pm \sigma_2^\prime \pm \sigma_{\mathrm{C}} \label{eq:thx2}\,,
\end{eqnarray}
where the uncorrelated uncertainty contributions are defined by:
\begin{eqnarray}
  \sigma_1^{\prime 2} & = & \sigma_1^2-\sigma_{\mathrm{C}}^2\,, \\
  \sigma_2^{\prime 2} & = & \sigma_2^2-\sigma_{\mathrm{C}}^2\,, \\
\end{eqnarray}
The BLUE combination in eq.~(\ref{eq:blue}) achieves an expression similar to a
regular weighted average:
\begin{equation}
  \hat{\theta} = \frac{\displaystyle
    \frac{\hat{\theta}_1}{\sigma_1^{\prime 2}} + 
    \frac{\hat{\theta}_2}{\sigma_2^{\prime 2}}
  }{\displaystyle
    \frac{1}{\sigma_1^{\prime 2}} + 
    \frac{1}{\sigma_2^{\prime 2}} 
    }\label{eq:bluew}\,,
\end{equation}
with weights that only take into account the uncorrelated uncertainty contributions,
but in this case the uncertainty receives an additional contribution due to
the correlated uncertainty, with respect to the uncertainty of the usual weighted average, and is given by:
\begin{equation}
  \sigma_{\hat{\theta}}^2 =\frac{1}{\displaystyle
    \frac{1}{\sigma_1^{\prime 2}} + 
    \frac{1}{\sigma_2^{\prime 2}} 
    } +\sigma_{\mathrm{C}}^2\,.
\end{equation}

Cases with negative weights have a less intuitive interpretation than eq.~(\ref{eq:bluew}),
as will be more evident in section~\ref{sec:neg}.

\section{Quantifying the importance of individual measurements}

In order to quantify the ``importance'' of each individual measurement used in a combination,
the first approach adopted in literature was to quote  the so-called ``relative importance'' (RI)
of each individual measurement,
proportional to the absolute value of the corresponding weight `$w_i$, and defined as:
\begin{equation}
  \mathrm{RI}_i = \frac{|w_i|}{\displaystyle\sum_{i=1}^n|w_i|}\,.
\end{equation}
The definition is chosen in order to have a normalization condition: $\sum_i \mathrm{RI}_i = 1$.
This approach was, for instance, used in combinations of top-quark mass measurements at Tevatron and
at LHC~\cite{cdfd01, atlascms1}.

This choice is questionable, as observed in ref.~\cite{Chierici}.
In fact, imagine we have three measurement, say $A$, $B_1$ and $B_2$.
The RI of measurement $A$ changes whether the three measurements are combined all together
or if $B_1$ and $B_2$ are first combined into $B$, and then $A$ and the partial
combination $B$ are combined together.

Ref.~\cite{Chierici}
proposes alternatives to RI based on 
the Fisher information, which is defined as:
\begin{equation}
  {\cal{J}}_{ij} = \left<
  \frac{\partial\ln L}{\partial\theta_i}
  \frac{\partial\ln L}{\partial\theta_j}
    \right>\,,
\end{equation}
the average being performed over all possible measurements, hence ${\cal J}$ does not
depend on a specific measurement, but only on the form of the likelihood function and
on the parameters choice.
Fisher information sets a lower bound to the variance of an unbiased
estimator~\cite{Cramer, Rao}:
\begin{equation}
  \sigma_{\hat{\theta}_i}^2 \ge {\cal{J}}_{ii}^{-1}\,,
\end{equation}
and for a single parameter, the Fisher information is given by:
\begin{equation}
  {\cal{J}} =   {\cal{J}}_{11} = {\vec{u}^{\mathrm{T}}\,\mathbf{V}\,\vec{u}} = \frac{1}{\sigma_{\hat{\theta}}^2}\,.
\end{equation}
The alternative quantities to RI proposed in ref.~\cite{Chierici} are the intrinsic information weight (IIW), defined as:
\begin{equation}
  \mathrm{IIW}_i = \frac{1/\sigma_i^2}{1/\sigma_{\hat{\theta}}^2} = \frac{1/\sigma_i^2}{\cal{J}}
\end{equation}
and the marginal information weight (MIW), defined as follows:
\begin{equation}
  \mathrm{MIW}_i = \frac{\Delta\cal{J}_i}{\cal{J}} = \frac{{\cal{J}} -
    {\cal{J}}_{\left\{ 1,\cdots, n \right\} - \{i\}}}{\cal{J}}\,,
\end{equation}
i.e.: the relative difference of the Fisher information of the combination and the Fisher information of the
combination excluding the $i^{\mathrm{th}}$ measurement.
Both IIW and MIW do not obey a normalization condition. For IIW the quantity
$\mathrm{IIW}_{\mathrm{corr}}$ can be defined such that:
\begin{equation}
  \sum_{i=1}^n \mathrm{IIW}_{i} + \mathrm{IIW}_{\mathrm{corr}} = 1\,.
\end{equation}
$\mathrm{IIW}_{\mathrm{corr}}$ represents the weight assigned to the correlation interplay,
not assignable to individual measurements,
and is given by:
\begin{equation}
  \mathrm{IIW}_{\mathrm{corr}} = \frac{1/\sigma_{\hat{\theta}^2} - \sum_{i=1}^n 1/\sigma_i^2}{1/\sigma_{\hat{\theta}}^2} =
  \frac{{\cal{J}} - \sum_{i=1}^n 1/\sigma_i^2}{\cal{J}}\,.
\end{equation}

Table~\ref{tab:tab} summarizes the properties of the different proposed
quantities.
\begin{table}[ht]
\centering
\caption{Properties of the different quantities proposed for the BLUE combination}
\label{tab:tab}       % Give a unique label
% For LaTeX tables you can use
\begin{tabular}{lcccc}
  \hline
  \multicolumn{2}{c}{Weight type} & $\ge 0$ & $\sum_i = 1$ &
  \begin{tabular}[x]{@{}c@{}}Consistent with \\partial combination\end{tabular} \\\hline
  BLUE coefficient & $w_i$ & \xmark & \cmark & \cmark \\
  Relative importance & $|w_i| / \sum_{i=1}^n |w_i|$ & \cmark & \cmark & \xmark \\
  Intrinsic Information Weight & $\mathrm{IIW}_i$ & \cmark & \xmark & \cmark \\
  Marginal Information Weight & $\mathrm{MIW}_i$ & \cmark & \xmark & \cmark \\
  \hline
\end{tabular}
% Or use
%\vspace*{5cm}  % with the correct table height
\end{table}

IIW and MIW are reported, instead of RI, in recent papers about combination of LHC and Tevatron measurements
related to top-quark physics~\cite{IIW1, IIW2, IIW3}.

\section{Negative weights}
\label{sec:neg}

\begin{figure}[ht]
% Use the relevant command for your figure-insertion program
% to insert the figure file.
\centering
\includegraphics[width=13cm,clip]{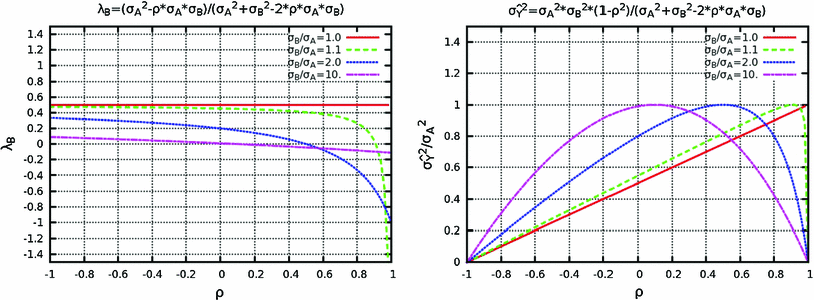}
\caption{
  BLUE coefficient for the second measurement $w_2$ (left; $\lambda_B$ in the original figure notation) and combined BLUE variance
  $\sigma_{\hat{\theta}}^2$  (right;  $\sigma_{\hat{Y}}^2$ in the original figure notation) as a function
  of the correlation $\rho$ between two measurements $1$ and $2$
  for various fixed values of the ratio $\sigma_2/\sigma_1$ ($\sigma_B/\sigma_A$ in the original figure notation).
  The figure is from ref.~\cite{Chierici}.}
\label{fig:chierici}
\end{figure}

Negative weights are always sign of a high-correlations regime.
The maximum value of the ratio $\sigma_{\hat{\theta}^2}/\sigma_1^2$ (eq.~(\ref{eq:blueerr})) as a function
of $\rho$ is obtained for $\rho = \sigma_1/\sigma_2$.
For $\rho > \sigma_1/\sigma_2$, an increase in correlation implies a decrease of the uncertainty
and a negative weight, and uncertainty becomes strongly dependent on $\rho$. 
For $\rho = \sigma_1/\sigma_2$, in particular, the weight $w_2$ becomes equal to zero,
as well as $\mathrm{MIW}_2 = 0$. But this does not imply that the measurement $\theta_2$
is not used in the combination.
Figure~\ref{fig:chierici}, from ref.~\cite{Chierici}, shows how the BLUE coefficient
$w_2$ and the ratio of uncertainties $\sigma_{\hat{\theta}}^2/\sigma_1^2$ vary as a function of the correlation $\rho$
for different fixed values of the ratio $\sigma_2/\sigma_1$ (note that the figure uses
a different notation with respect to this text, as specified in the caption).

\begin{figure}[ht]
% Use the relevant command for your figure-insertion program
% to insert the figure file.
\centering
\includegraphics[width=7cm,clip]{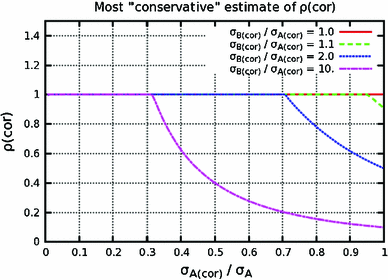}
\caption{The
  most ``conservative'' value of an unknown correlation $\rho$(cor) between $\sigma_A$(cor) and $\sigma_A$(cor)
  as a function of $\sigma_A$(cor)$/\sigma_A$, for several values of $\sigma_B$(cor)$/\sigma_A$(cor)$\ge 1$.
  The figure is from ref.~\cite{Chierici}.}
\label{fig:chierici2}
\end{figure}

When the correlation coefficient $\rho$ is not well know, 
assuming $\rho=1$ is not always a conservative choice.
The assumption $\rho=1$ yields the largest possible uncertainty
only if the uncorrelated contributions to the total uncertainty dominate.
$\rho$ should be accurately determined in case of negative weights in order to avoid the risk
of underestimating uncertainties.
Assume that the two measurements $A$ and $B$ have total uncertainties given by the sum in quadrature
of uncorrelated contributions, $\sigma_A$(unc) and $\sigma_B$(unc) respectively,
and correlated contributions, $\sigma_A$(cor) and $\sigma_B$(cor) respectively, whose
correlation coefficient is $\rho$(cor).
Figure~\ref{fig:chierici2} shows the most ``conservative'' value of the correlation
coefficient $\rho$(cor), which is equal to 1 only for $\sigma_B$(cor)$/\sigma_A$(cor)$<(\sigma_A/\sigma_A$(cor))$^2$.

\section{Bias with the BLUE method}

The BLUE method provides an unbiased estimate only if uncertainties and their correlation are known exactly.
This is not always a realistic scenario, since estimates of uncertainties and their correlation, and not
their true values, are available in most of the cases.
Moreover, the uncertainty estimates may depend on the assumed central value.
One example of such a case in which the BLUE method provides a bias is the combination of two
Poissonian estimates, each of which has an uncertainty estimate that depends on the central value
through a square-root relation:
\begin{eqnarray}
  \hat{n}_1 & = & \hat{n}_1 \pm \sqrt{\hat{n}_1}\,,\label{eq:pois1}\\
  \hat{n}_2 & = & \hat{n}_2 \pm \sqrt{\hat{n}_2}\,.\label{eq:pois2}
\end{eqnarray}
A maximum-likelihood estimate, using Poissonian distributions, would produce the following
unbiased combined estimate:
\begin{equation}
  \hat{n} = \frac{1}{2}\left(
  \hat{n}_1+\hat{n}_2 \pm \sqrt{\hat{n}_1 + \hat{n}_2}
  \right)\,.
\label{eq:poisml}
\end{equation}
The BLUE combination, instead, gives weights proportional to $1/\hat{n}_i$, which correspond to
a harmonic average:
\begin{equation}
  \hat{n} = \frac{2\hat{n}_1\hat{n}_2}{\hat{n}_1+\hat{n}_2}\pm\sqrt{\frac{\hat{n}_1\hat{n}_2}{\hat{n}_1+\hat{n}_2}}\,.
\label{eq:poisbl}
\end{equation}
Equation~(\ref{eq:poisbl}), compared to eq.~(\ref{eq:poisml}), exhibits a bias
because, due to the dependence of uncertainties on the measured values,
downward measurement fluctuations achieve larger weights pulling down the combination,
while upward fluctuations produce a smaller opposite effect.

In many cases, relative uncertainty estimates are available, like for uncertainty contributions
due to luminosity, efficiencies, etc.
When performing a combination, the best estimate of the central value improves the individual
measurements, hence one may argue whether
the assumed uncertainties should change accordingly,
re-evaluating them using the central-value
estimate from the BLUE combination~\cite{Lyons, Lista}. The method can be applied
iteratively until the combination converges to a stable value. Namely:
\begin{eqnarray}\
  \hat{\sigma}_i^{(0)} = \hat{\sigma}_i(\hat{\theta}_i)\,,\,\,
  \hat{\rho}_{ij}^{(0)} = \hat{\rho}(\hat{\theta}_i, \hat{\theta}_j) & \xrightarrow{\,\,\,\mathrm{BLUE}\,\,\,} & \hat{\theta}^{(1)}\\
  \hat{\sigma}_i^{(1)} = \hat{\sigma}_i(\hat{\theta}^{(1)})\,,\,\,
  \hat{\rho}_{ij}^{(1)} = \hat{\rho}(\hat{\theta}^{(1)}, \hat{\theta}^{(1)}) & \xrightarrow{\,\,\,\mathrm{BLUE}\,\,\,} & \hat{\theta}^{(2)}\\
  \hat{\sigma}_i^{(2)} = \hat{\sigma}_i(\hat{\theta}^{(2)})\,,\,\,
  \hat{\rho}_{ij}^{(2)} = \hat{\rho}(\hat{\theta}^{(2)}, \hat{\theta}^{(2)}) & \xrightarrow{\,\,\,\mathrm{BLUE}\,\,\,} & \hat{\theta}^{(3)}\\
  & \cdots &\nonumber
\end{eqnarray}
In practice, convergence only needs few iterations in most of the cases.

Let us assume that a contribution to the total uncertainty is known as relative uncertainty.
In this case, uncertainties can be written as the sum in quadrature of a contribution that does not depend on the
central value and another contribution that is proportional to the central value\footnote{
  This was not the case for the combination of Poissonian measurements from eq.~(\ref{eq:pois1}), (\ref{eq:pois2}),
  where the dependence on central value was not linear, but it is a convenient assumption in many realistic cases.
}:
\begin{eqnarray}
  \theta & = & \hat{\theta}_1 \pm \sigma_1 \pm r_1 \hat{\theta}_1\,, \\
  \theta & = & \hat{\theta}_2 \pm \sigma_2 \pm r_2 \hat{\theta}_2\,.
\end{eqnarray}
The covariance matrix estimate can be written as follows:
\begin{equation}
  \hat{\mathbf{V}} = \left(
  \begin{array}{cc}
    \sigma_1^2 +(r_1\hat{\theta}_1)^2 & \rho_0\sigma_1\sigma_2+\rho_r r_1 r_2 \hat{\theta}_1\hat{\theta}_2 \\
    \rho_0\sigma_1\sigma_2+\rho_r r_1 r_2 \hat{\theta}_1\hat{\theta}_2  & \sigma_2^2+(r_2\hat{\theta}_2)^2
  \end{array}
  \right)
\end{equation}
where $\rho_0$ and $\rho_r$ represent the correlation coefficients of the uncertainty
contributions that do not depend on the central value and of the ones that are proportional to the central value, respectively.

A special case is when uncertainties are fully proportional to the central values, i.e.:
$\sigma_1= 0$ and $\sigma_2 = 0$. In that case, the iterative BLUE method converges in
two iterations to the following central value:
\begin{equation}
  \hat{\theta} = \frac{
    (r_2^2 -\rho_rr_1r_2) \hat{\theta}_1 + (r_1^2-\rho_r r_1 r_2) \hat{\theta}_2
  }
  {q_1^2 -2\rho_r r_1 r_2 + r_2^2}\,,
\end{equation}
which is similar to eq.~(\ref{eq:blue}) but relative uncertainties are used
in place of the absolute ones.

\begin{figure}[htp]
% Use the relevant command for your figure-insertion program
% to insert the figure file.
\centering
\includegraphics[width=6cm,clip]{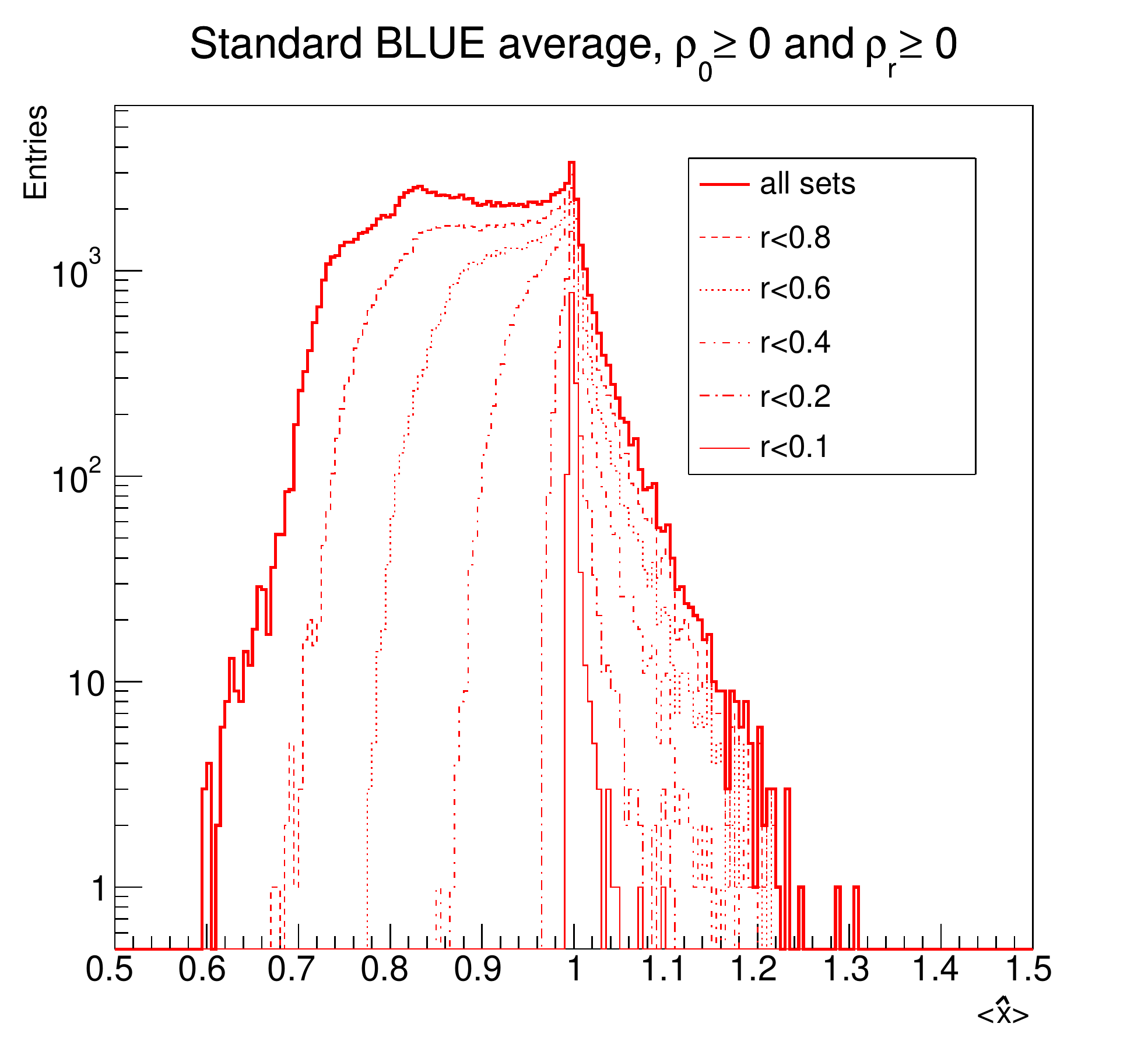}
\includegraphics[width=6cm,clip]{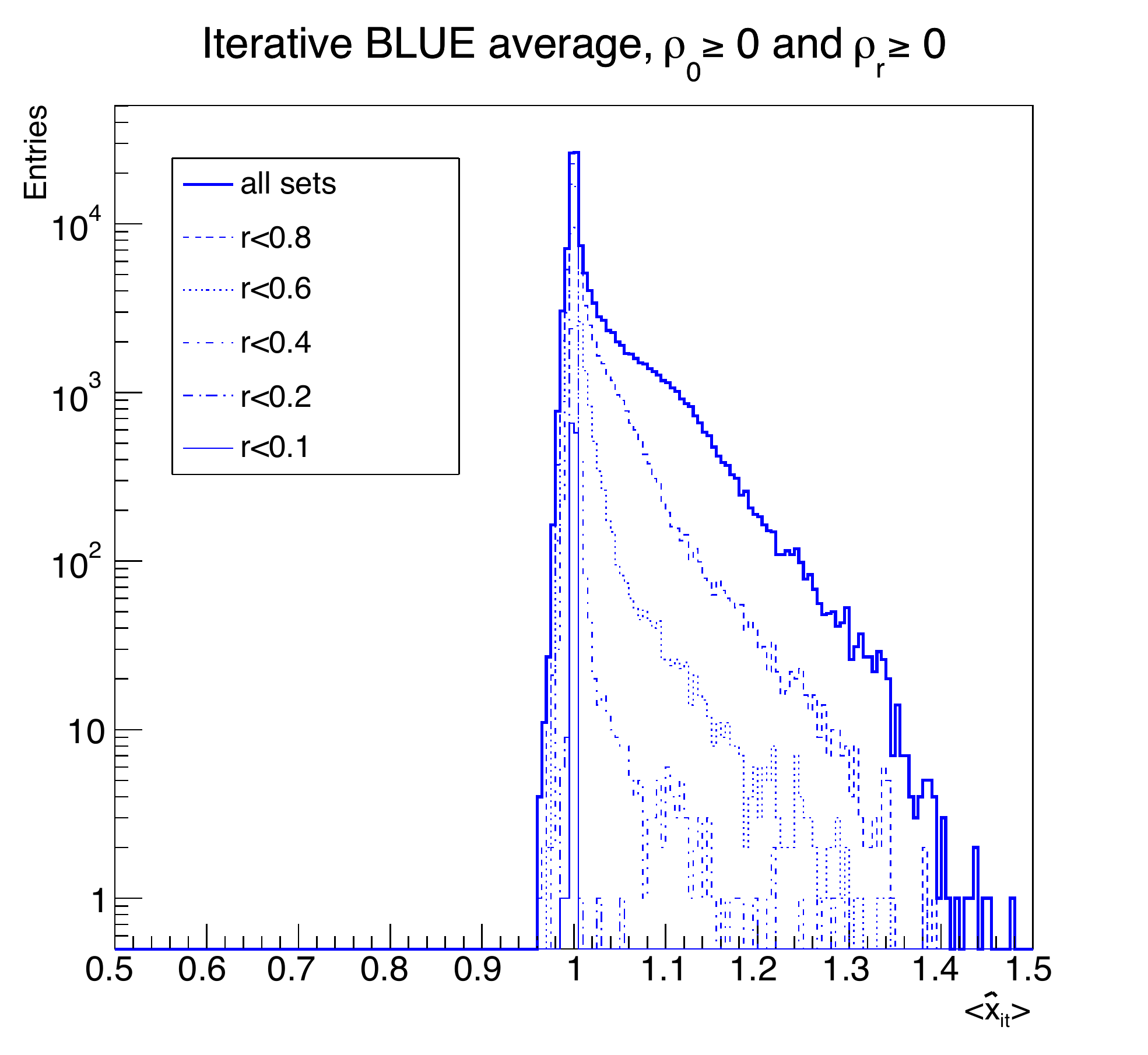}
\includegraphics[width=6cm,clip]{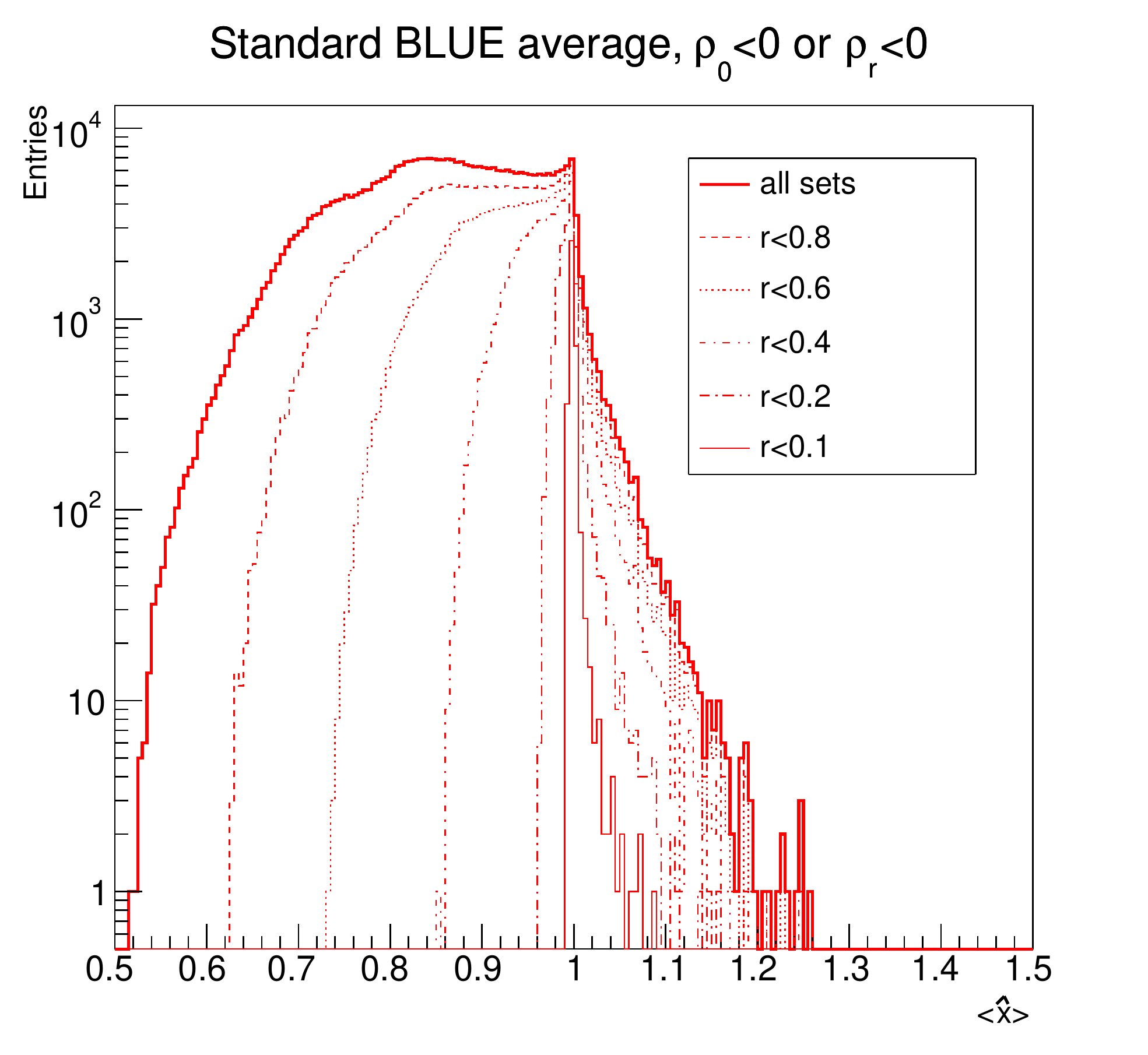}
\includegraphics[width=6cm,clip]{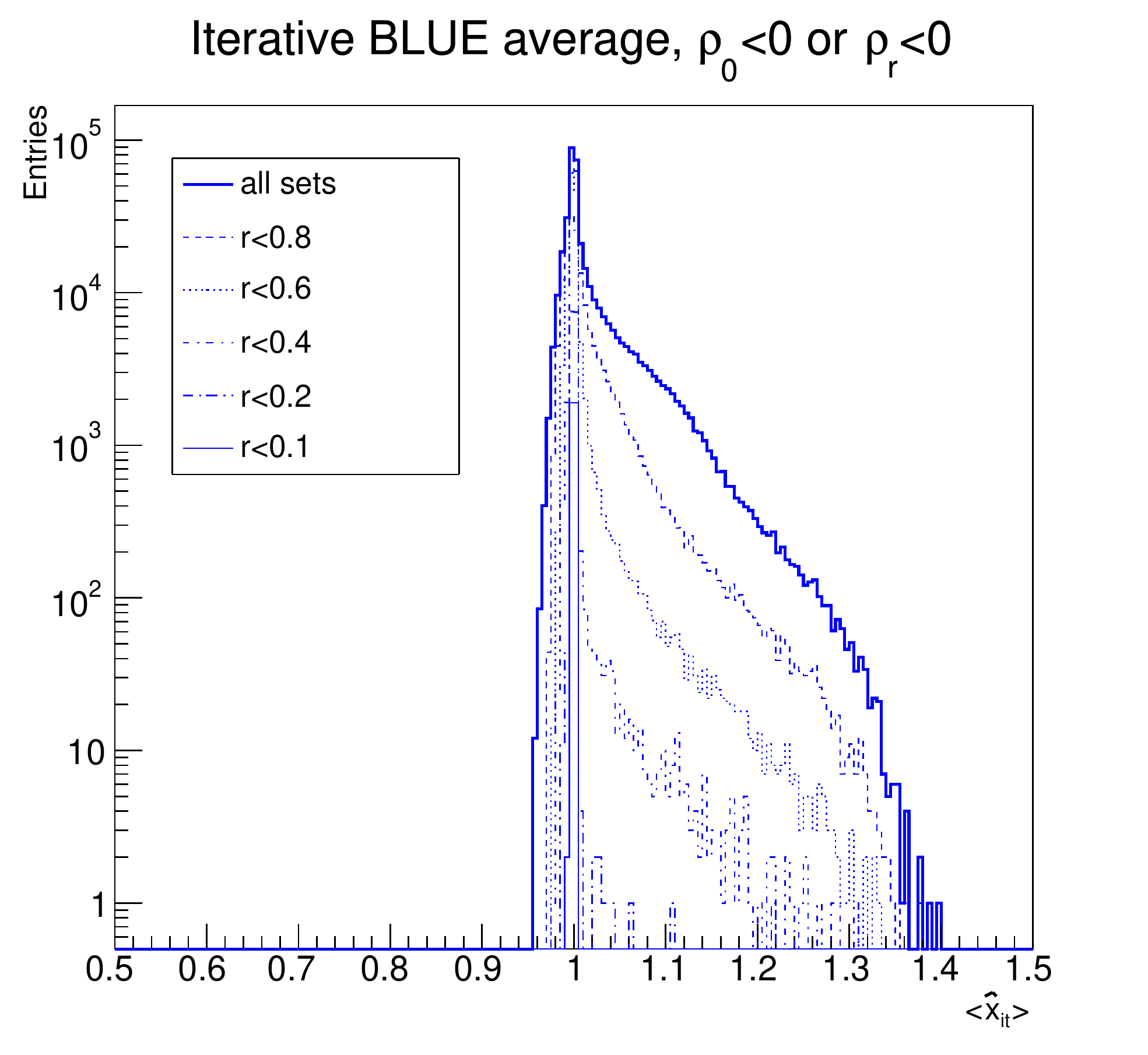}
\includegraphics[width=6cm,clip]{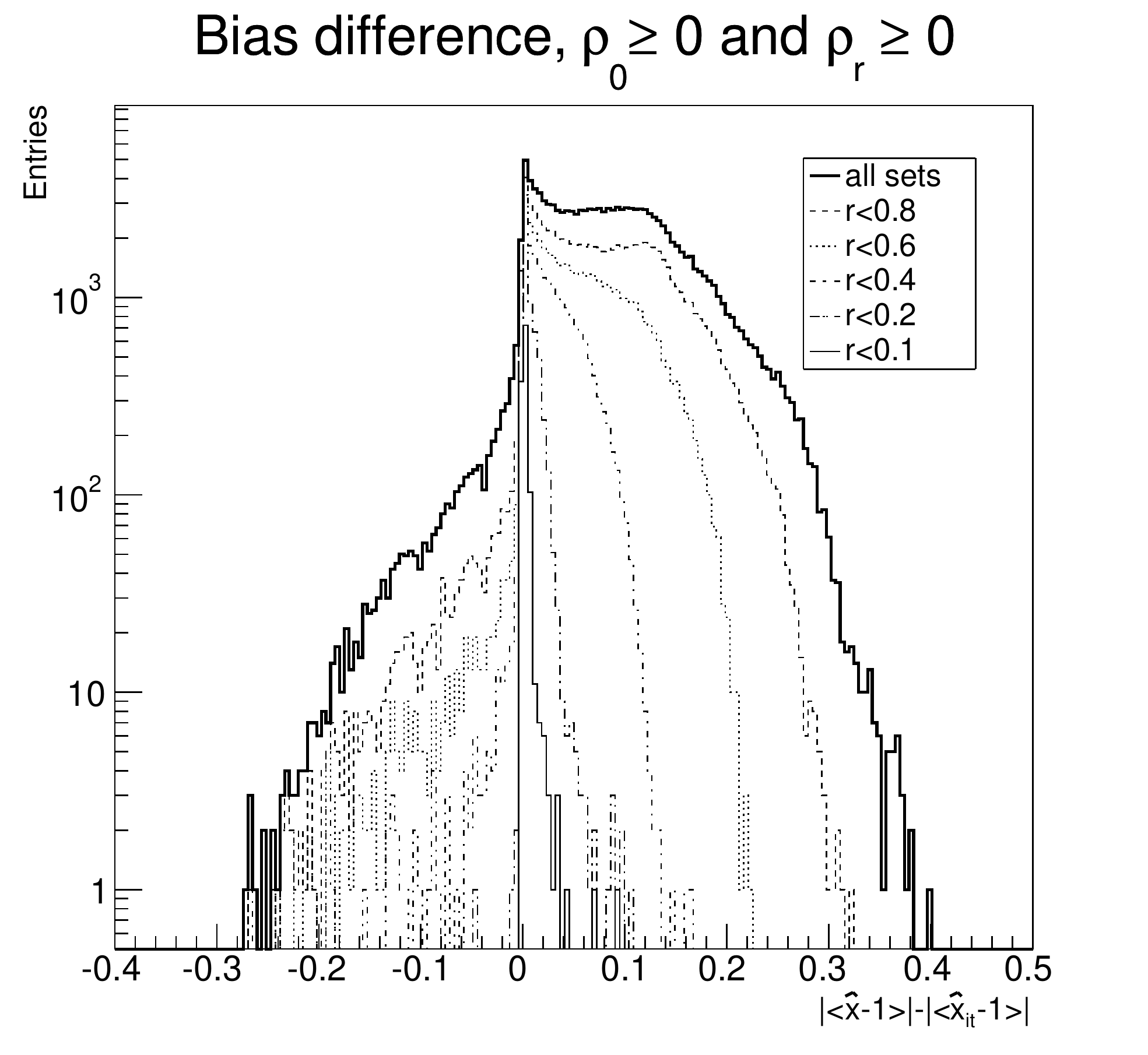}
\includegraphics[width=6cm,clip]{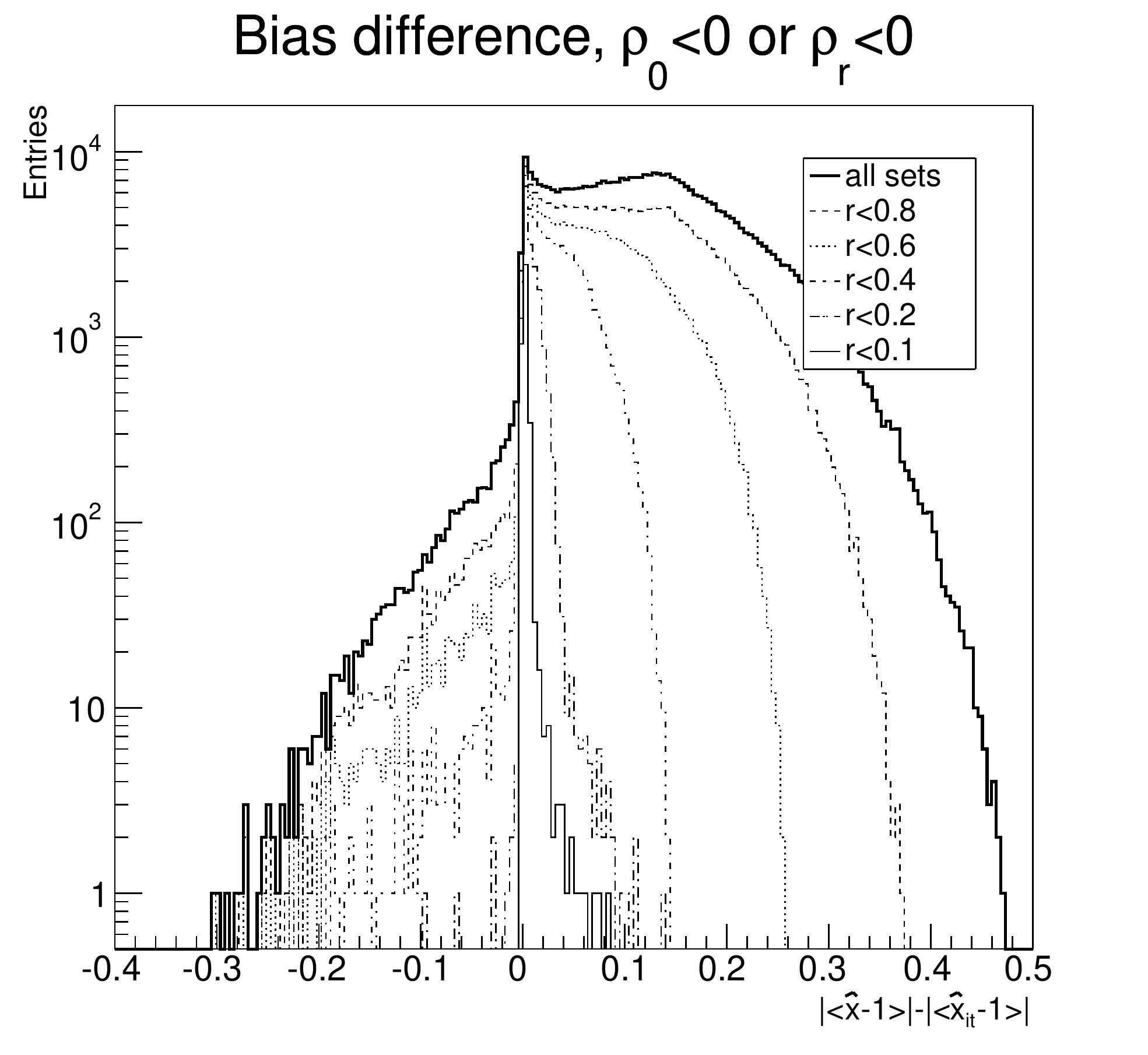}
\caption{Distribution of the average value of the
  standard (top and middle, left) and iterative (top and middle, right) BLUE estimates
  for different limits on $r_{1,\,2}$ and for $\rho_0\ge 0$ and $\rho_r \ge 0$ (top)
  and for $\rho_0< 0$ and $\rho_r < 0$ (middle).
  Bottom plots show the difference of measured absolute value of the bias for the
  standard and iterative BLUE estimates for different limits on $r_{1,\,2}$ and for
  $\rho_0\ge 0$ and $\rho_r \ge 0$ (left)
  and for $\rho_0< 0$ and $\rho_r < 0$ (right).
  Positive values indicate a smaller bias in the iterative method compared to the standard method.
  Underestimate of the uncertainty (negative values) occur in fewer cases with the iterative BLUE method
  than with the standard BLUE method.
  The figure is from ref.~\cite{Lista}.
  }
\label{fig:lista}
\end{figure}

\begin{figure}[htp]
% Use the relevant command for your figure-insertion program
% to insert the figure file.
\centering
\includegraphics[width=6cm,clip]{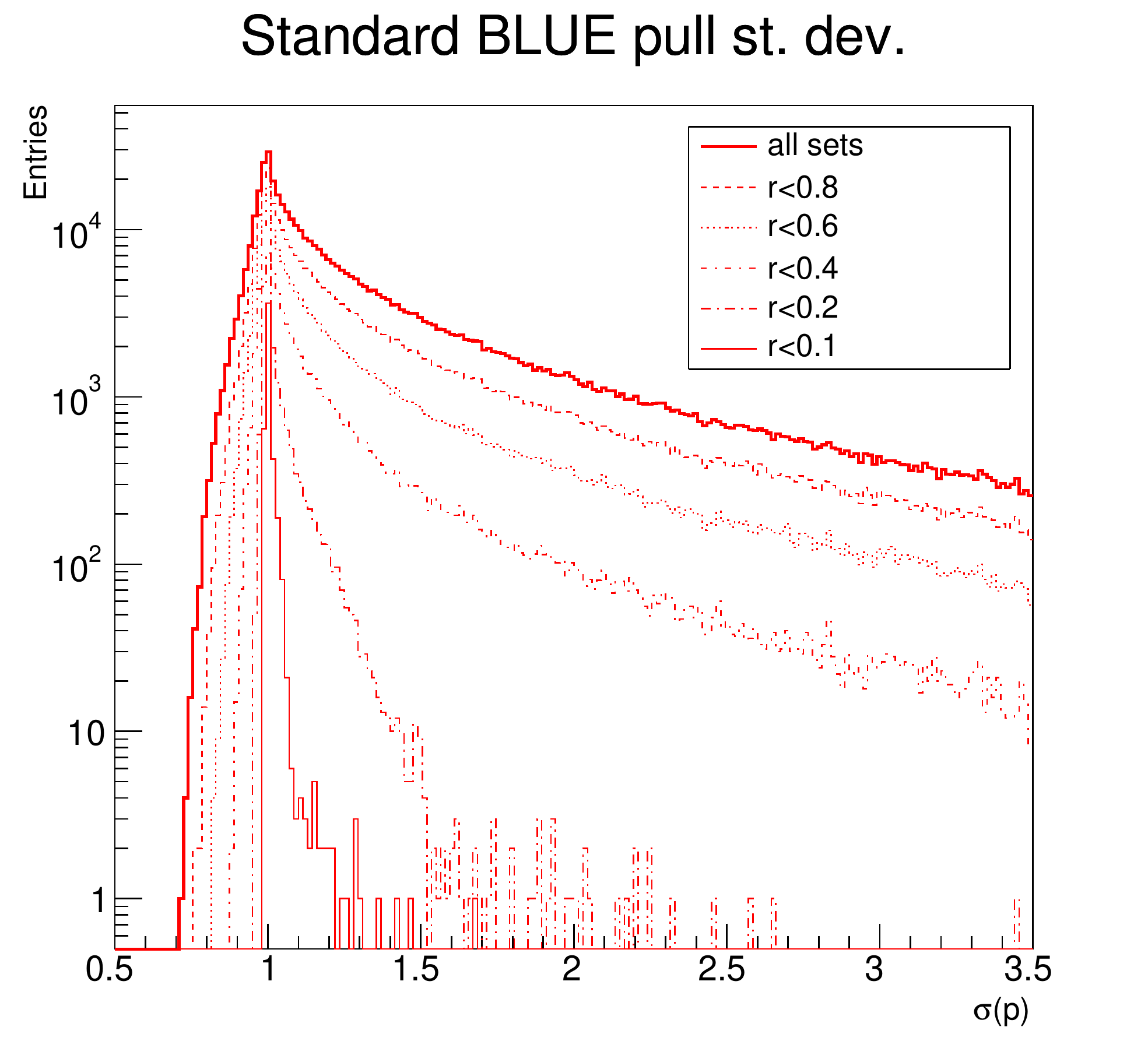}
\includegraphics[width=6cm,clip]{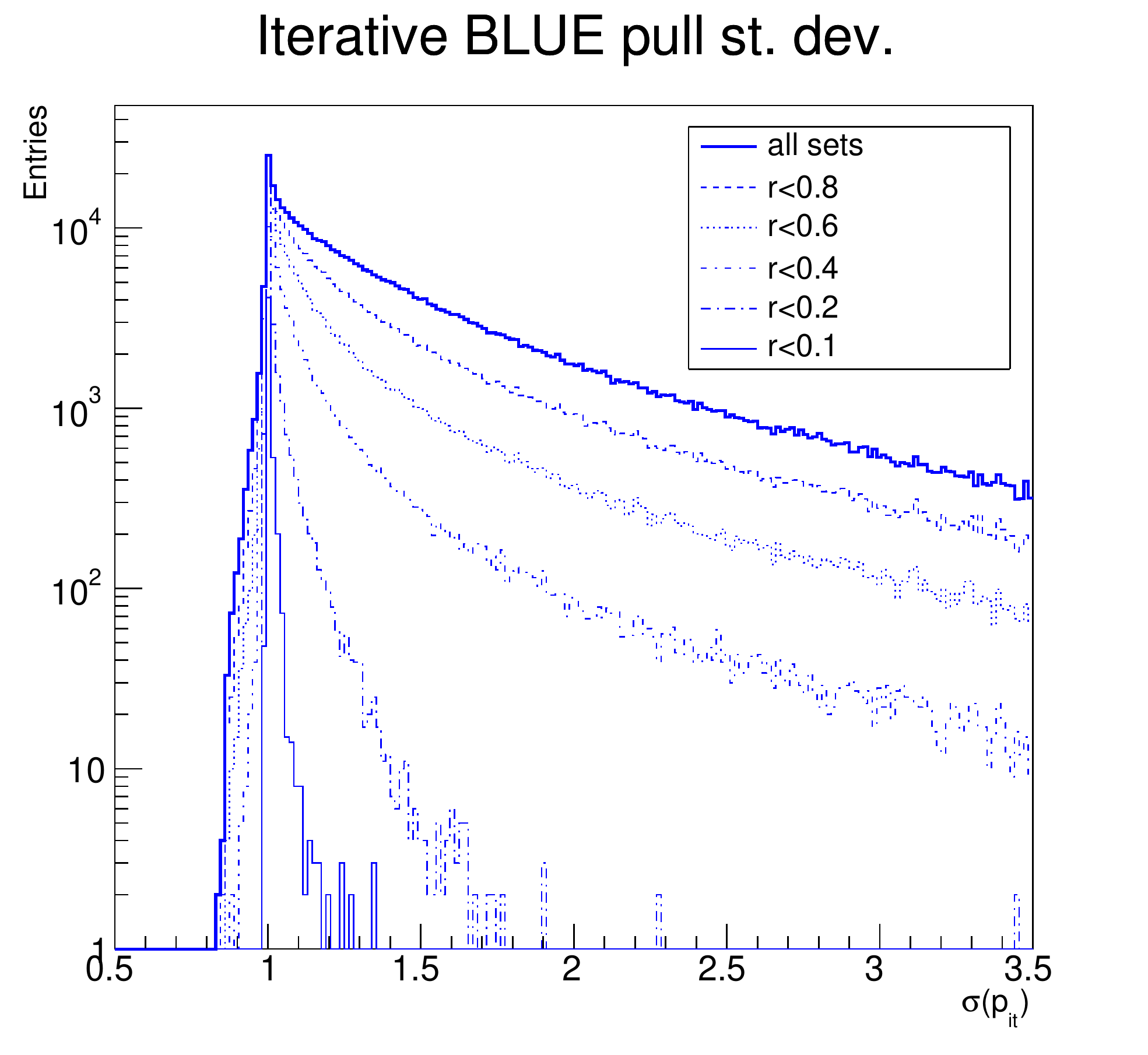}
\caption{Distribution of the average value of the pull standard deviation for the standard (left) and iterative (right) BLUE estimates for different limits on $r_{1,\,2}$.
 The figure is from ref.~\cite{Lista}.
  }
\label{fig:lista2}
\end{figure}

A numerical Monte Carlo study~\cite{Lista} shows that in most of the cases the bias in the combination
can be mitigated by applying the iterative procedure.
Assuming for simplicity, and without loss of generality, a true value $\theta=1$,
uncertainties and their correlations are chosen randomly by spanning a wide
ranges of values in the boundaries $\sigma_1,\, \sigma_2,\,r_1\,,r_2\,, < 1$ and $-1 < \rho_0,\,\rho_r < 1$.
For each set of randomly-extracted uncertainties and correlation values, 500\,000 random values of
$\hat{\theta}_1$ and $\hat{\theta}_2$ are generated using the proper
two-dimensional correlated Gaussian distribution.
The BLUE method is then applied in its standard formulation and iteratively.
Pulls distributions allow to study the bias and the correctness of uncertainty estimates.
Bias is in general mitigated with the iterative BLUE method while
the standard BLUE method tends to underestimate the central value,
as shown in Fig.~\ref{fig:lista}.
The iterative BLUE application may provide overestimates
in few of the cases with large uncertainties.
More detailed comparisons of the two method are available in ref.~\cite{Lista}.
In particular, the combination uncertainty may be overestimated
(and in fewer cases underestimated) in both standard and iterative BLUE,
but the iterative BLUE estimates tend to be more conservative,
as visible in Fig.~\ref{fig:lista2}.
A dedicated toy Monte Carlo may be useful in case of large individual uncertainties
in order to achieve a proper combined uncertainty estimate.

\section{Conclusions}
Combination of measurements is a crucial task to improve the precision on the knowledge of important
parameters.
Combining individual likelihood functions before applying any statistical method
is the most rigorous and precise approach, but often the full likelihood function of individual measurements
is not available.
When uncertainties and their correlations are available, the BLUE method is a simple and
powerful tool to combine measurements. But 
BLUE may have counterintuitive behaviors such as negative weights or uncertainties
that may decrease for increasing correlation.
Correlation needs to be accurately evaluated in those cases, and an assumption of a
100\% correlation is not necessary a conservative choice.
In case relative uncertainty contributions are known, or more in general when
uncertainty estimates depend on the central values, the BLUE method may exhibit a bias that can be mitigated,
in most of the cases, with an iterative application of the method where
uncertainties are rescaled at each iteration to the combined central value.

\end{document}